\documentclass{article}

\usepackage{arxiv}
\usepackage{authblk}
\usepackage[table,dvipsnames]{xcolor}
\usepackage[utf8]{inputenc} 
\usepackage[T1]{fontenc}    
\usepackage{hyperref}       
\usepackage{url}            
\usepackage{booktabs}       
\usepackage{amsfonts}       
\usepackage{nicefrac}       
\usepackage{microtype}      
\usepackage{cleveref}       
\usepackage{lipsum}         
\usepackage{graphicx}
\usepackage{doi}
\usepackage{multirow}
\usepackage{tikz}
\usepackage{hyperref}
\definecolor{lightgray}{rgb}{0.83, 0.83, 0.83}

\definecolor{very_important}{rgb}{0.5, 0.5, 0.5}
\definecolor{important}{rgb}{0.6, 0.6, 0.6}
\definecolor{rather_important}{rgb}{0.7, 0.7, 0.7}
\definecolor{not_important}{rgb}{0.8, 0.8, 0.8}

\usepackage{longtable}
\usepackage{array}
\usepackage[utf8]{inputenc}
\usepackage{booktabs}
\usepackage[natbib=true, backend=biber, bibstyle=numeric, sorting=none]{biblatex}
\newcolumntype{P}[1]{>{\raggedright\arraybackslash}p{#1}}
\newcolumntype{C}[1]{>{\centering\arraybackslash}p{#1}}
\newcommand*\circled[1]{\tikz[baseline=(char.base)]{
            \node[shape=circle,draw,inner sep=1pt] (char) {#1};}}

\title{Tailoring Stakeholder Interests to Task-Oriented Functional Requirements}

\date{}

\AtEveryBibitem{%
    \ifentrytype{article}{%
        \clearfield{url}%
        \clearfield{urldate}%
        \clearfield{month}%
        \clearfield{notes}%
        \clearfield{issn}%
    }{}%
    \ifentrytype{inproceedings}{%
        \clearfield{url}%
        \clearfield{urldate}%
        \clearfield{month}%
        \clearfield{notes}%
        \clearfield{issn}%
    }{}%
}

\DeclareSourcemap{
  \maps[datatype=bibtex]{
    \map{
      \step[fieldset=issn, null]
      \step[fieldset=month, null]
      \step[fieldset=urldate, null]
    }
  }
}

\author[1,2]{Philipp Haindl}
\author[2]{Reinhold Plösch}
\affil[1]{Software Competence Center Hagenberg, Austria}
\affil[2]{Department of Business Informatics - Software 
Engineering, Johannes Kepler University Linz, Austria}

\renewcommand{\headeright}{}
\renewcommand{\undertitle}{Technical Report}
\renewcommand{\shorttitle}{Tailoring Stakeholder Interests to Task-Oriented Functional Requirements}

\hypersetup{
pdftitle={A template for the arxiv style},
pdfsubject={q-bio.NC, q-bio.QM},
pdfauthor={Philipp Haindl, Reinhold Plösch}
}

\begin{document}

\maketitle

\begin{abstract}
Without a specific functional context, non-functional requirements can only be approached as cross-cutting concerns and treated uniformly across all features of an application. This neglects, however, the heterogeneity of non-functional requirements that arises from stakeholder interests and the distinct functional scopes of software systems, which mutually influence how these non-functional requirements have to be satisfied. Earlier studies showed that the different types and objectives of non-functional requirements result in either vague or unbalanced specification of non-functional requirements. We propose a task analytic approach for eliciting and modeling user tasks to approach the stakeholders' pursued interests towards the software product. Stakeholder interests are structurally related to user tasks and each interest can be specified individually as a constraint of a specific user task. These constraints support DevOps teams with important guidance on how the interest of the stakeholder can be satisfied in the software lifecycle sufficiently. We propose a structured approach, intertwining task-oriented functional requirements with non-functional stakeholder interests to specify constraints on the level of user tasks. We also present results of a case study with domain experts, which reveals that our task modeling and interest-tailoring method increases the comprehensibility of non-functional requirements as well as their impact on the functional requirements, i.e., the users' tasks.
\end{abstract}


\section*{Remark}
This paper is an addition to our peer-reviewed publications in \cite{haindl_se_workhsop,haindl_research_2019} and contains a more elaborate presentation of our findings. When citing our work, please always cite the peer-reviewed publications; citing this technical report should always be optional for cases where you explicitly reference aspects that were not published in the peer-reviewed publications.

\section{Introduction}
Intertwining functional and non-functional requirements (NFRs) is a challenging endeavor in software projects of any scale. The primarily engineering-oriented understanding of NFRs neglects the practical relevance of  business, strategic, operational, legal and privacy interests which must be taken into account throughout the DevOps cycle. We use the term \textit{stakeholder interest} to emphasize this broader understanding of NFRs. In industry, these stakeholder interests are typically specified as cross-cutting concerns throughout the application without considering relevance, applicability, and characteristic of an interest for a certain software feature. As the complexity of satisfying an interest differs between features, this lack of preciseness in specifications results in undetected non-functional dependencies between components and features in development, as well as increasing operational efforts throughout the DevOps cycle \cite{zowghi_requirements_2005}. Focusing on concrete tasks that users will perform with a software system allows functional units of work to be elicited from a user perspective. In contrast to isolated features, tasks can be seen as functional units that focus on the users' goals and therefore required interactions between features. This makes the individual relevance of competing NFRs for a single user task more tangible. 
In our work, we understand constraints as refinements and operationalizations of NFRs. Constraints of a user task can unambiguously be validated for fulfillment. Especially among non-technical stakeholders, this facilitates requirements negotiation \cite{blaine_software_2008} and assessment of tradeoffs in satisfying constraints \cite{berander_requirements_2005}. 

In this paper we present an approach for eliciting and modeling user tasks based on a modified form of hierarchical task analysis \cite{annett_hierarchical_2003}. It facilitates the tailoring of stakeholder interests to user tasks so that the interests can be refined into constraints and satisficed during the full software lifecycle. The prime objective of these constraints is to \textit{satisfice} the original interest, i.e., to \textit{satisfy} an interest \textit{sufficiently} and not better than required. The precise refinement of each interest as a constraint on the level of user tasks also gives DevOps teams a more tangible understanding of how an individual constraint must be practically handled in software design, implementation, and operation.

Our approach augments the computer-implemented development method outlined by US Pat. No. 15/661,498 \cite{prod_dev_patent}, which aims to find suitable quantitative constraints for NFRs through Monte-Carlo simulations. Our work explicitly facilitates elicitating, refining and specifying constraints for individual functional requirements with quantitative measures. Also, it supports requirements engineers in the process of evaluating possible trade-offs of contradicting NFRs on the level of user tasks. 

The remainder of this paper is organized as follows. An overview of related work and the demarcation of our approach to other works are given in section \ref{related_work}. In section \ref{research_context} we describe the research context and study design we chose to answer the research questions. Following, section \ref{tacos_approach} introduces our approach to eliciting, modeling and tailoring functional requirements and stakeholder interests. Section \ref{evaluation} presents an evaluation of this approach with experts from software engineering, operations, and product management before we describe the possible threats to validity in section \ref{threats}. The conclusion of our work and possible directions for future work are presented in section \ref{conclusion}.

\section{Related Work}
\label{related_work}
Goal- \cite{lamsweerde_goal-oriented_2001,rolland_modeling_2005} and hence task-driven \cite{lauesen_task_2012,zowghi_requirements_2005} approaches have proven to be effective for requirements elicitation. They do, however, not guide stakeholders to specify suitable NFRs for a certain product or service. According to Fotrousi et al. \cite{fotrousi_quality_2014}, the key limitation of goal models is that the impact of an unsatisfied NFR towards a goal is difficult to understand for stakeholders. Riegel et al. \cite{riegel_systematic_2015} elaborated on a prioritization method which classifies non-engineering related NFRs by project-related, financial, customer, operational business performance or business-strategy related benefits.
Karlsson et al. \cite{karlsson_requirements_2007} investigated the applicability of goal models in the context of market-driven software development and highlighted the benefits of goal-centered feature elicitation. They showed that these models facilitate stakeholder participation in the requirements engineering process.

Regnell et al. \cite{regnell_supporting_2008} presented the QUPER approach for relating NFRs to development costs needed for their fulfillment. According to their observations, stakeholders lack a shared understanding of the required efforts, and thus costs, for fulfilling NFRs in practice. Their framework QUPER allows to prioritize and balance NFRs through comparison with products of competitors.
Cleland-Huang et al. \cite{cleland-huang_goal-centric_2005} presented a method for tracing quality goals of different levels of granularity in specification documents, offering a means to systematically review interrelations between quality goals and system components. Hermann et al. \cite{herrmann_exploring_2007} compare two methods for refining NFRs from quality goals. Again, the authors stress the importance of specifying NFRs individually for functional requirements and also to scrutinize the relation between NFRs and business goals.

Another stream of research tackles the different satisfaction criteria of NFRs in the context of individual user tasks and software features. Zubcoff et al. \cite{zubcoff_evaluating_2019} propose a Pareto-based approach for specifying soft goals in the context of user tasks which assists requirements engineers in evaluating trade-offs between NFRs. The authors also underline that NFRs shall be specified individually for functional requirements to improve end-user satisfaction. This argumentation is also taken by Ameller et al. \cite{ameller_dealing_2010} who compare approaches for model-driven design in software engineering. The authors conclude that most model-driven design approaches do not intertwine functional requirements and NFRs which may result in the expectations of the stakeholders not being fully satisfied. Ameller et al. \cite{ameller_dealing_2019} present a survey among practitioners in 18 companies about the practice of NFR specification in model-driven design. The interviewees argued that NFRs are not only difficult to specify through models, but even difficult to discover and explicate in measurable terms. As a result of this imprecise definition, the fulfillment of NFRs usually can only be evaluated very late in the software development process. Also, the intervieews reported that in their companies unmet NFRs in projects often require retrofitting of functional requirements. An interview study of Svensson et al. \cite{berntsson_svensson_quality_2009} also shows that in practice, NFRs have a lower priority than functional requirements and are partly not even specified in the early stages of development. This lack of integration between functional requirements and NFRs can result in long time-to-market and cost overruns in software projects \cite{chung_non-functional_2009,chung_non-functional_2000,doerr_non-functional_2005,daneva_software_2013}.

In summary, recent research had elaborated the importance of balancing functional and NFRs in understanding technical qualities, but no structured method has yet been presented that captures interests outside the technical domain. As can be recognized, several authors especially underlined the suitability of goal-centered approaches to increase stakeholder participation and facilitate shared understanding. The approach presented in this paper explicitly links these two strands of research, providing a structured method for eliciting and specifying constraints from NFRs for individual functional requirements. Our approach also permits stakeholders to have different domains and thus interpretations of NFRs before gradually refining and eventually collectively specifying appropriate constraints therefor with quantitative measures.

\section{Research Context and Study Design}
\label{research_context}
\begin{figure*}[t]
\centering
\includegraphics[trim=0cm 0cm 0cm 0cm, scale=0.9]{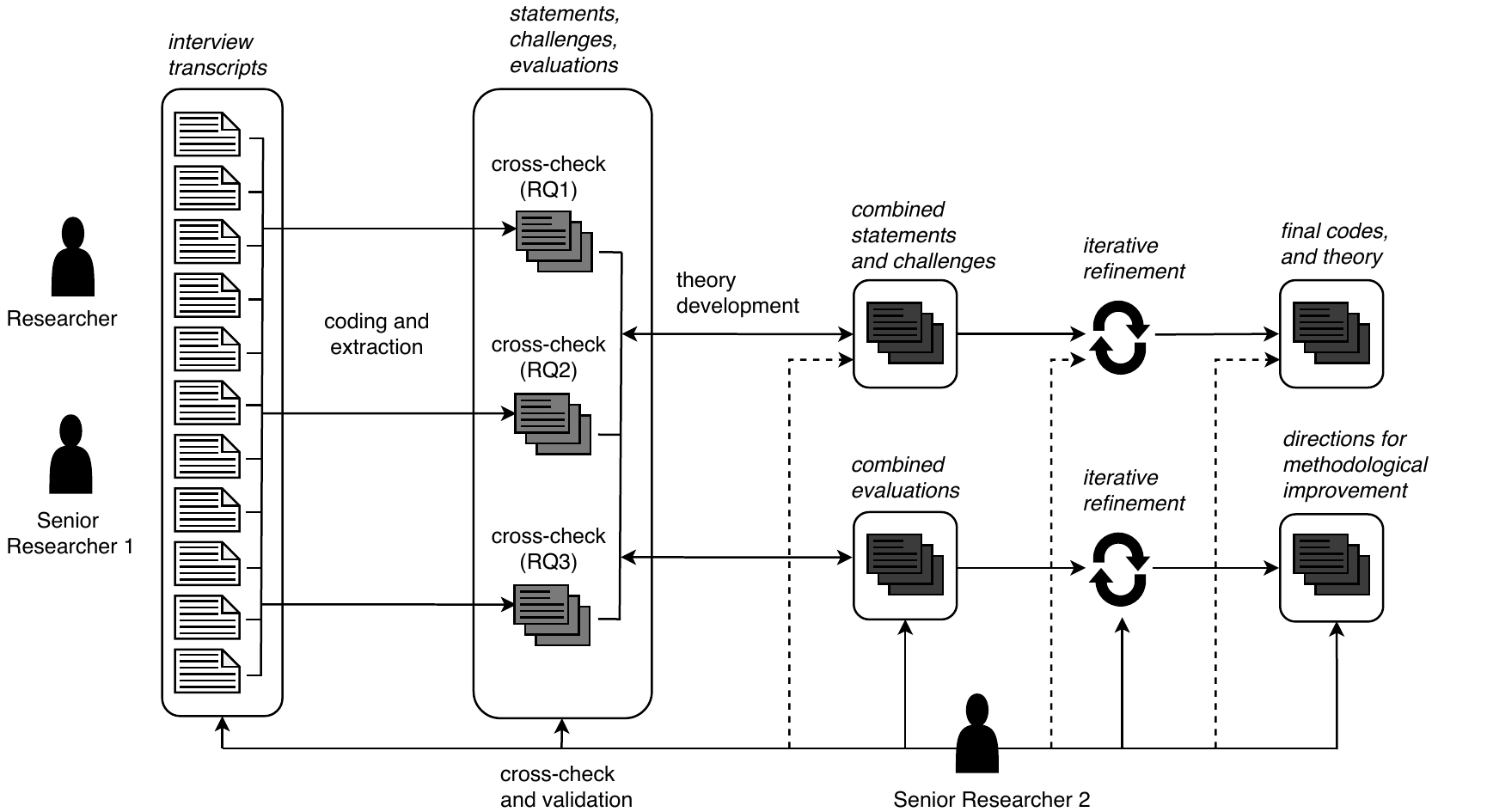}
\caption{Research methods for iteratively deriving codes, a theory, and further improvement directions for our task-oriented interest tailoring approach through expert interviews (adapted from \cite{vierhauser_case_2014}).\label{fig:research_method}}
\end{figure*}

Our study started with a literature review of key challenges for eliciting and specifying NFRs using moderation-based prioritization techniques for assessing individual relevance for software features. Originating from informal discussions we expanded this literature review by human task analysis to assess how these methods can be used to elicit and prioritize functional requirements from user tasks. 

Based on this literature review we designed an exploratory case study using expert interviews to research the elicitation practice and specification preciseness of NFRs in different companies. Following, we presented the experts our approach to the task-oriented modeling of functional requirements and tailoring stakeholder interests to elicited user tasks. 
\subsection{Case Study Design}
Taking the guidelines by Runeson and H{\"o}st \cite{runeson_guidelines_2009} as a blueprint we designed the questionnaire covering practices, challenges, and problems that arise during specifying NFRs with different stakeholders. Also, we conducted a pilot interview as suggested by Yin \cite{yin_case_2017} with one highly experienced expert and included his feedback to improve the questionnaire itself. Particulary, we refined certain phrases in the questionnaire and used more common terminology to assure a good understanding during the interviews among the experts.

The questionnaire comprised 20 open questions and 4 closed questions on a 4-point Likert scale and was separated into 3 parts: The first part captured educational and company background, roles in projects and years of experience with requirements engineering. In the second part we asked questions to find out how the companies currently model functional requirements and NFRs, what challenges they are confronted with, and what types of stakeholder interests typically need to be taken into account thereby. Finally, in the third part we presented the experts our approach for tailoring stakeholder interests to concrete constraints in the context of user tasks. During the interviews we used figure \ref{fig:flow} to present the approach to the experts. In this final part we asked them to rate suitability and applicability of the approach as well as anticipated benefits and drawbacks.

We conducted 11 face-to-face expert interviews with senior software engineers, technical product and project managers, and requirements engineers from different companies in Austria. The interviewees' companies have a mean employee count of 65 and ranged from small enterprises to companies with more than 1000 employees. Industry sectors comprise public administration, finance, logistics, IT consultancy, public media, ERP software development, and software product manufacturing. The experts have an average of 13 years professional experience with requirements engineering, and each expert typically held multiple roles in the projects. The primary roles of the experts were software development and architecture (28\%), business (14\%) or requirements management (14\%), project coordination (14\%), and technical product (9\%) and infrastructure management (9\%). This distribution of roles reflects the interweaving between functional, quality, and resource aspects in software projects.
\subsection{Research Questions}
The exploratory case study should help us to answer 3 research questions:
\begin{itemize}
	\item{\textbf{RQ1 - What elicitation and specification practices for NFRs are used in industry?} This question focuses on the preciseness and granularity of elicited requirements and how interdependencies between contradicting requirements are detected.}
	\item{\textbf{RQ2 - What are typical classes of stakeholder interests in software projects?} This question intends to clarify how the different types of NFRs can be attributed to certain types of stakeholders (e.g., from legal, business or engineering domains).}
	\item{\textbf{RQ3 - How do requirement engineers rate our approach for modeling and tailoring stakeholder interests in the context of user tasks?} The focus of this question is to gather feedback from experts particularly regarding the proposed modeling approach for functional requirements and the methodology for tailoring stakeholder interests to constraints.}
\end{itemize}
\subsection{Data Analysis Procedure}
The interviews took between 50 and 90 minutes and were performed and transcribed by one researcher. Subsequently, the interview transcripts were analyzed by two researchers and the answers to the closed questions were coded quantitatively. The answers to the open questions where coded in the transcripts for the later extraction of qualitative data. At that point we did not categorize the statements to ensure that we do not exclude potentially important qualitative data at an early stage of analysis (cf. figure \ref{fig:research_method}).

Following, the qualitative statements were categorized conjointly by two researchers using a grounded theory \cite{adolph_using_2011,glaser_discovery_2000} approach. The objective hereby was to assign each statement meaningfully to a category and later draw practical conclusions from the answers. The categories were developed iteratively and similar statements combined in one category until consensus was reached among the two researchers. A second senior researcher cross-checked the process but was not involved in transcribing or coding the interviews.

From the 11 interviews we extracted a total of 763 statements (68 on average per interview). After several iterations we condensed 59 categories out of the  qualitative statements expressed in the interviews. Finally, we identified 9 classes of stakeholder interests that best categorize the different objectives pursued through NFRs. The final classification schema for stakeholder interests is discussed in section \ref{interests_classification_schema} and our aggregated findings from analyzing the interviews are presented in section \ref{evaluation}.

\section{Description of the Approach}
\label{tacos_approach}

\begin{figure*}[h]
\centering
\includegraphics[trim=0cm 0cm 0cm 0cm, scale=0.74]{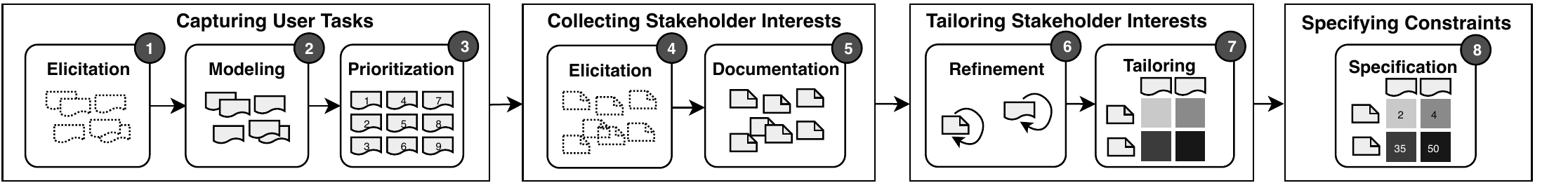}
\caption{The method defines a structured sequence from capturing user tasks to collecting and tailoring stakeholder interests for the specification of task-dependent constraints.\label{fig:process}}
\end{figure*}

The described approach hierarchically decomposes functional blocks of software into tasks which can be executed by the user. In contrast to eliciting isolated features, this gives a more comprehensive picture about how these different functional blocks are related to each other so that the user can achieve the goal of each task with the software. To facilitate integration and participation of stakeholders from multiple domains, each step of the method is repeated until all stakeholders have a sufficient understanding of tasks and user goals. Figure \ref{fig:process} illustrates the eight steps of the approach, starting with capturing user tasks and collecting interests. 

\subsection{Capturing User Tasks}
\noindent \circled{1} \textbf{Elicitation:}
To elicit tasks, we rely on hierarchical task analysis \cite{annett_hierarchical_2003} with some modifications to carve out the functional scope of the elicited actions. Hierarchical task analysis breaks down a task into goals, subgoals, plans, and operations, focusing on the structure and decomposition of the task into a hierarchy of subtasks in sequential order of tasks so that the goal can be attained. According to this notion, a \textit{task} is a sequence of actions that people perform to attain a goal. The elicitation of tasks in our method follows a sequential procedure that provides structured guidance to detail the functional scope and connections between tasks.
\begin{itemize}
\item[--] \textbf{Step 1: Define task under analysis.} Which task should be analyzed and what is its objective? Scope boundaries must be clearly defined prior to analysis, typically driven by the underlying business model.
\item[--] \textbf{Step 2: Collect data for task analysis.} This collection comprises data about task execution, dependencies between tasks, constraints, and interviewing subject matter experts or key users.
\item[--] \textbf{Step 3: Determine the overall task goal.} This will become the root goal of the hierarchy, i.e., the starting point for decomposition.
\item[--] \textbf{Step 4: Determine subtasks.} The predecessor task goal will then be used to derive subtasks necessary for achieving the superordinate goal. 
\item[--] \textbf{Step 5: Identify task details.} Derived from the goal of each subtask, identify the intentions of the user in attaining the goal and the resulting responsibilities of the system to support these intentions. We elaborate this step in more detail below.
\item[--] \textbf{Step 6: Define execution plans.} Execution plans organize how to reach the goal of the task by modeling execution order and dependencies between the subtasks. The objective is to define how the subtasks relate to each other so that the goal of the task can be achieved.
\end{itemize}

While hierarchical task analysis allows infinite refinement of tasks to the point that tasks are purely operational for a user, our method only allows refinement of tasks by means of \textit{task details} tables. This forces the development team to define the granularity of  tasks upfront. If the refinement is still too vague to be operational, a separate model should be created for refinement. This assures that tasks and constraints can be properly treated and described at the refined level.

\begin{figure*}[h]
\centering
\includegraphics[trim=0cm 0cm 0cm 0cm, scale=0.8]{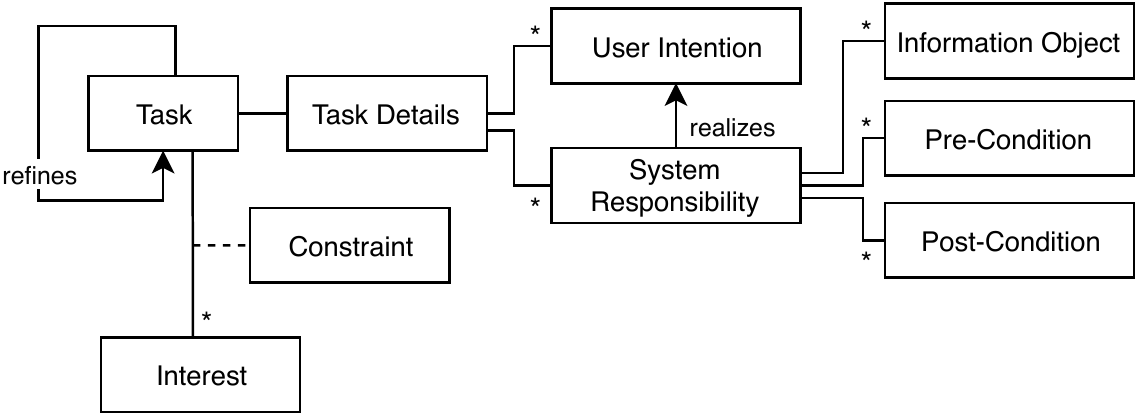}
\caption{Meta-model of the approach for specifying  task-dependent stakeholder interests.}\label{fig:meta}
\end{figure*}

\begin{figure*}[h]
\centering
\includegraphics[trim=0cm 0cm 0cm 0cm, scale=0.6]{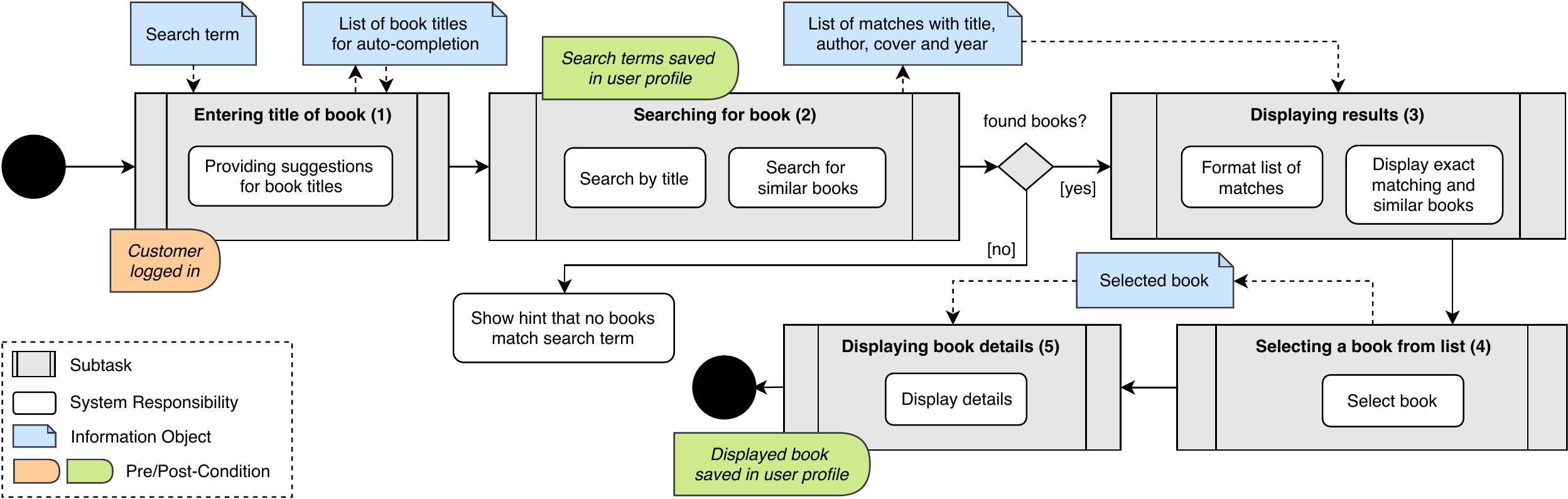}
\caption{Modeling Subtasks, System Responsibilities, Information Objects and Conditional Flows.}\label{fig:flow}
\end{figure*}

For the structured elicitation of task details, our approach offers two perspectives: (1) \textit{user intentions}, the user's interactions during execution of the task; and (2) \textit{system responsibilities} to support these user intentions through the system. Both perspectives are compared with each other in tabular form and refined with \textit{pre-} and \textit{postconditions}, as well as \textit{information objects}, which describe the information generated or required for task execution. The relations between these concepts are defined in the meta-model illustrated in figure \ref{fig:meta}. The task detailing step helps to map the fine-grained intentions of the user to suitable functional requirements by carving out the minimal and satisfactory technical solution  to execute the task effectively. \newline

\noindent \circled{2}~\textbf{Modeling:}
In the next step, the control flow and decision points between the subtasks are modeled. Figure~\ref{fig:flow} shows an exemplary task of searching for a book at an online store. This example was also used for presenting our approach in the expert interviews. Each subtask is modeled as a rectangular gray box labeled with the corresponding \textit{user intention} and incorporates one or multiple rounded rectangles (i.e., \textit{system responsibilities}) that reflect the required software functionality. Information objects are denoted by colored rectangles, with the direction of the arrows indicating whether the information object is generated or consumed by the respective subtask. The objective of this modeling is to outline the control flow among the subtasks for all stakeholders and to facilitate sketching the required software features. In this work we refer to features as concrete software implementations \cite{bourque_guide_2014} which realize the actions that are required to implement a task.
 Also, the granularity of the actions executed within a subtask can be chosen arbitrarily to foster a common understanding of the scope and goal of the features among the involved stakeholders. In contrast to process modeling, this type of modeling primarily focuses on the flow among the activities and also relates them to the technical counterparts needed for their execution. For the sake of simplicity, we abstain to describe conditional loops or concurrent executions of subtasks in our modeling approach.\newline
\\
\circled{3} \textbf{Prioritization:}
Next, only those user tasks are selected for the subsequent steps, which are most important (following the idea of a minimum viable product (MVP)), bear a competitive advantage, or have been selected by an agile team for development in the next program increment. Typically, the question of importance can only be answered by considering the underlying business model. Also, the required assessment of the tasks' value contribution prevents specifying details or exhaustive constraints for rarely executed tasks.
\subsection{Collecting Stakeholder Interests}
\noindent\circled{4}~\textbf{Elicitation:}
\label{interest_elicitation}
Due to their primary technical focus, the notion of NFRs does not satisfactorily cover non-technical objectives of stakeholders having an impact on a software system. This specially comprises all objectives that must be considered continuously throughout the software lifecycle, from software development and operation to its decommissioning. Existing classification schemes \cite{glinz_rethinking_2005,broy_rethinking_2015} for these non-technical objectives, which we understand as \textit{stakeholder interests}, also reflect their relevance for software quality and need for precise specification. These stakeholder interests are still unsatisfactorily covered by existing, solely technically oriented standards such as ISO/IEC 29148:2011 \cite{noauthor_iso/iec/ieee_2011-1} or ISO/IEC 25030:2007 \cite{noauthor_iso/iec/ieee_2018}, which often results in NFRs being only vaguely elicited.

The objective of this step is to capture all stakeholders' quality-related, operational, business, legal, and other non-behavioral interests that influence how the later software system needs to support the tasks of the users. We extended existing classification schemes by additional interests, which we gathered from the expert interviews. This interest classification schema practically aids the elicitation of the interests and is presented in section \ref{interests_classification_schema}.\newline \newline 

\begin{table*}[t]
\caption{Tailoring stakeholder interests: The results from qualitatively evaluating individual relevance of interests for each user task (\textit{task-interest matrix}, left) allow to derive concrete quantitative constraints (\textit{task-constraint matrix}, right).}\label{tab:task_constraint_matrix}
\vspace{10pt}
\hspace{-20pt}
  \begin{minipage}{.65\linewidth}
    \begin{tabular}{m{1pt}|p{54pt}|m{35pt}|m{35pt}|m{35pt}|m{35pt}|m{20pt}} 
     \multicolumn{2}{c}{} & \multicolumn{4}{c}{\textbf{User Tasks}} & \\
      \cline{3-6}
      \multicolumn{2}{c|}{} & Search \newline for book & Update \newline credit card \newline information & Change  shipping  address & Write \newline book review & \\
      \cline{2-6}
      \multirow{20}{*}{\rotatebox[origin=c]{90}{\textbf{Stakeholder Interests}}}
      & The software \newline must be \newline responsive to \newline user inputs. & \cellcolor{very_important}  & \cellcolor{important} & \cellcolor{important} & \cellcolor{not_important} & \\ \cline{2-6}  
      & The software\newline must handle\newline peaks of con-\newline current users. & \cellcolor{very_important}  & \cellcolor{very_important} & \cellcolor{rather_important} & \cellcolor{important} & \\ \cline{2-6}  
      & The software\newline must be \newline maintainable\newline effortlessly. & \cellcolor{very_important}  & \cellcolor{important} & \cellcolor{important} & \cellcolor{not_important} & \begin{tikzpicture}
\draw[->, line width=0.4mm] (0,0) -- (0.5,0); \end{tikzpicture}\\ \cline{2-6}
      & The software shall be \newline modularized \newline into separate \newline units. & \cellcolor{not_important}  & \cellcolor{very_important} & \cellcolor{very_important} & \cellcolor{not_important} & \\ \cline{2-6}  
      & The software\newline must be \newline resilient to \newline external \newline outages. & \cellcolor{very_important}  & \cellcolor{very_important} & \cellcolor{important} & \cellcolor{rather_important} & \\ \cline{2-6}  
      \end{tabular}
    \end{minipage}    
    \begin{minipage}{.20\linewidth}
      \centering
    \begin{tabular}{|m{35pt}|m{35pt}|m{35pt}|m{35pt}|} 
      \multicolumn{4}{c}{\hspace{15pt}\textbf{User Tasks}} \\
      \cline{1-4}
       Search \newline for book & Update \newline credit card \newline information & Change \newline shipping \newline address & Write \newline book review \\ \cline{1-4}
      \cellcolor{very_important} \vspace{5pt} response time $<$ 2 \newline ms  \vspace{5pt} & \cellcolor{important}  response time $<$ 3 \newline ms & \cellcolor{important} response time $<$ 3 \newline ms & \cellcolor{not_important} response time $<$ 4 \newline ms \\ \cline{1-4}
       \cellcolor{very_important} \vspace{5pt} $>$ 100 \newline users per\newline minute \vspace{5pt} & \cellcolor{very_important} $>$ 100 \newline users per\newline minute & \cellcolor{rather_important} $>$ 50 \newline users per\newline minute & $>$ 70 \newline users per\newline minute \cellcolor{important}  \\ \cline{1-4}
       \cellcolor{very_important} \vspace{5pt} technical debt \newline $<$ 1 day \vspace{5pt} & \cellcolor{important} technical debt \newline $<$ 2 days & 
      \cellcolor{important} technical debt \newline $<$ 2 days & \cellcolor{not_important} technical debt \newline $<$ 3 days\\ \cline{1-4}
        \cellcolor{not_important} \vspace{8pt} class fan \newline out $<$ 10 \vspace{8pt} & \cellcolor{very_important} class fan \newline out $<$ 4 & \cellcolor{very_important} class fan \newline out $<$ 4 & \cellcolor{not_important} class fan \newline out $<$ 10 \\ 
\cline{1-4}
       \cellcolor{very_important} \vspace{9pt} MTBF \newline $<$ 1 min \vspace{9pt} & \cellcolor{very_important}  MTBF \newline $<$ 1 min & \cellcolor{important}  MTBF \newline $<$ 5 min & \cellcolor{rather_important} MTBF \newline $<$ 10 min\\ \cline{1-4}
    \end{tabular}
\end{minipage}%
\end{table*}

\noindent \circled{5} \textbf{Documentation:}
 In this step, the elicited stakeholder interests are documented informally to express the stakeholders' objectives and expectations towards the software system. It is important to document each stakeholder interest in a manner comprehensible for all involved stakeholders to foster shared understanding and also to later assess its individual relevance for other interests. In this step of the method an interest does not need to be documented on a quantitative basis. The concrete tailoring and deriving of constraints for each user tasks is done in a subsequent step. Stakeholder interests can be documented in the following ways:
\begin{itemize}
\item[$\bullet$]  ``The software must be responsive to user inputs.''
\item[$\bullet$]  ``... handle peaks of concurrent users.''
\item[$\bullet$]  ``... be deployable automatically.''
\item[$\bullet$]  ``... recover quickly after outages.''
\item[$\bullet$]  ``... be operated in the cloud.''
\item[$\bullet$]  ``In-Transit customer data must be encrypted.''
\end{itemize}

Our method explicitly separates interest elicitation from documentation. The former being conducted by all stakeholders in a brainstorming like manner striving to unveil as many possibly relevant interests that could impact the software lifecycle; the latter just to document the results from the elicitation step.

\subsection{Tailoring Stakeholder Interests}
\noindent \circled{6} \textbf{Refinement:}
Similar to the refinement of user tasks, also stakeholder interests are iteratively refined until they show a delimited and comprehendible scope for the later tailoring. Refined stakeholder interests again are documented conjointly with the relevant stakeholders. This is done to cross-check that there is common understanding about the interest even after its refinement. Stakeholder interests which are too large in scope are decomposed up to a suitable level. The overall objective of this step is to prevent ambiguities about an interest's scope among the stakeholders. Uniform comprehension about the scope of an interest is the prerequiste for its effective tailoring and the specification of suitable constraints therefore.\newline

\noindent \circled{7} \textbf{Tailoring:}
In this step, the stakeholder interests and user tasks are analyzed pairwise to evaluate the relevance of an interest and how it can be satisfied in each narrow task context. This detailed analysis also assures that all elicited stakeholder interests eventually are made operational for software engineers at the level of individual user tasks.

Stakeholder interests and user tasks are then related to each other in a two-dimensional \textit{task-interest matrix}. Following, each interest is tailored individually to each user task so that it can be fulfilled exactly for the respective user task. Resulting from this tailoring is a qualitative ordering of the cells indicating the relevance of the interest for the respective user task, e.g., through coloring cells darker gradually with relevance. 

This qualitative ordering is shown on the left side of table \ref{tab:task_constraint_matrix}. It also shows that at that point of the method no quantitative fulfillment criteria for the stakeholder interests are defined. The focus of this step is on getting a common understanding about the relevance of an interest for a particular user task. As a practical example, the interest that \textit{``the software must be responsive to user inputs''} might be subjected to different qualitative expectations depending on the concrete user task. Responsiveness might be much more important for a user when searching for a book than when writing a book review, but less important when changing the shipping address. As an example from the perspective of the software manufacturer, the interest \textit{``the software shall be modularized into separate units.''} is  important when changing credit card information or the shipping address of the customer. These two tasks are part of the billing process and thus the associated source code need to be better modularized than for other tasks. 

Analyzing these interests outside the context of individual user tasks neglects the fact that some tasks are more important than others and thus need more attention. The fulfillment of interests typically requires intertwining multiple teams, e.g., requirements engineering and DevOps teams. This even more emphasizes the need to evaluate its relevance conjointly with all relevant stakeholders in the narrow context of a user task. Also, stakeholder interests evaluated with little relevance for multiple user tasks should be revised as this indicates ambiguities regarding its scope or its actual irrelevance.

\subsection{Specifying Constraints}
\noindent \circled{8} \textbf{Specification:}
In the last step of our method, the matrix created in the last step serves as input for  eventually specifying concrete constraints. These constraints are specified individually for each user task to \textit{satisfice} the stakeholder interests. Hence, when specifying concrete constraints the objective is to fulfill the stakeholder interests \textit{satisfactorily and not optimally}, so that a majority of stakeholder interests can be fulfilled.

The previous qualitative evaluation of individual relevance of an interest for a user task helps to find concrete metrics and criteria for specifying the constraints. Ideally, one should start with determining metrics and values for the most relevant user task for a stakeholder interest. Taking the example illustrated in table \ref{tab:task_constraint_matrix}, the task to search for a book needs to be specially responsive. Determining the response time as metric and measurements of below 2 milliseconds as being very responsive, the other constraints can be derived thereof straightforwardly. This procedure is repeated until
\begin{itemize} 
  \item for each cell in the matrix there is a concrete constraint defined, or alternatively
  \item the stakeholders agree the interest is of little relevance for a certain user task and does not need to be specified.
\end{itemize}

Table \ref{tab:task_constraint_matrix} shows the transition from the qualitative evaluation of the interests' relevance (left) into concrete constraints for each user task (right). The main outcome of this step is a 2-dimensional \textit{task-constraint} matrix incorporating user tasks, stakeholder interests and constraints consolidated between the involved stakeholders.

\section{Case Study Results}
\label{evaluation}
In this section we present selected results that we extracted from the interview transcripts. The concrete research context and design of the case study is described in section \ref{research_context}. We group our findings correspondingly to the research questions. Verbal statements from the interviews are italized and reflect the given answer of the expert without our interpretation.

\subsection{Practice of Eliciting and Specifying Non-Functional Requirements}
\label{evaluation_rq1}
The first research question addresses the elicitation and specification practices for NFRs in the interviewees' companies. Hereby we carve out the current state of practice in the companies to derive possible directions for improvements through our approach. We asked the experts 3 questions to find out \begin{itemize}
\item which methods they use to elicit NFRs, and
\item how precise they elicit certain categories of NFRs, and
\item how they detect contradicting NFRs during requirements engineering.
\end{itemize}

\begin{figure}[h]
\centering
\includegraphics[scale=0.6]{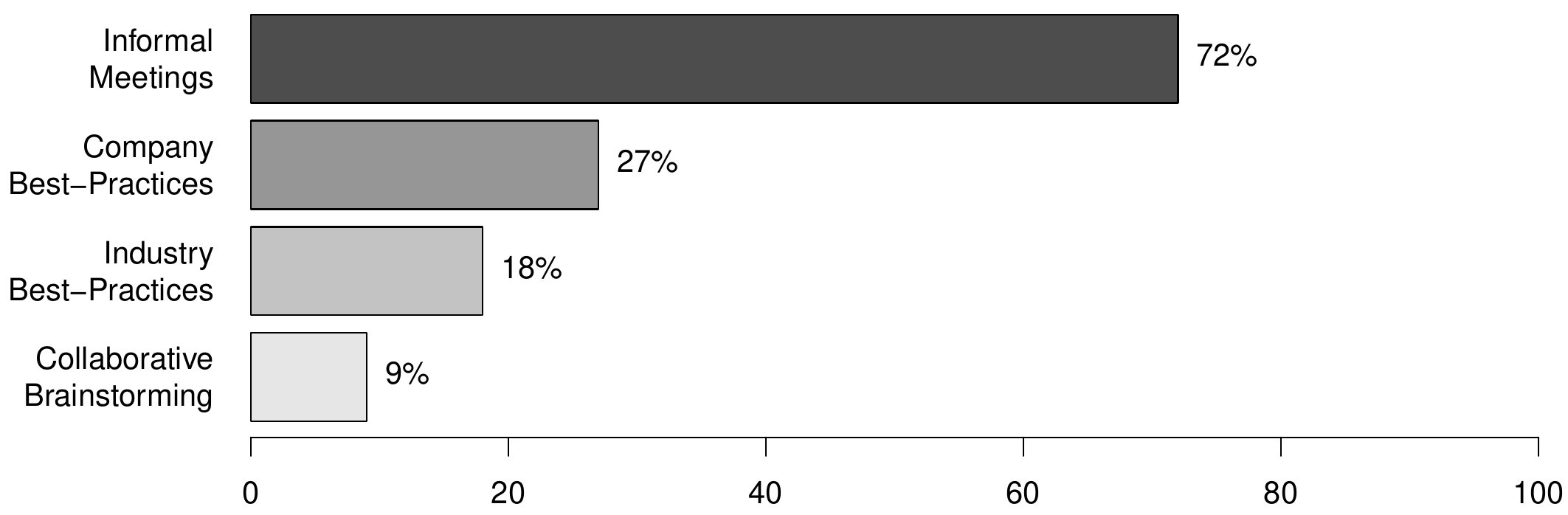}
\caption{Practice of NFR elicitation in the interviewed companies.\label{fig:nfr_elicitation_method}}
\label{fig:nfr_elicitation}

\end{figure}

As shown in figure \ref{fig:nfr_elicitation}, in 72\% of the companies, NFRs are elicited through informal meetings, i.e., without any structured method or checklist. The respective experts explained that their companies usually abstain from eliciting NFRs during requirements engineering. This is because their companies are convinced that \textit{``...the engineers know what it should work like''} and because they \textit{``...have always done it that way''}. In 27\% of the companies, NFRs are elicited based on predefined templates of best-practices. While these templates do not specify concrete constraints or measures how to evaluate the fulfillment of NFRs, they provide blueprints how to generically handle them during software development. Industry-best practices, i.e., commercial templates for generic specification of NFRs are used in 18\% of the interviewed companies. Only in 9\% of companies, NFRs are elicited collaboratively in a brainstorming like manner and tailored to an individual software project. 

\begin{figure}[h]
\centering
\includegraphics[scale=0.6]{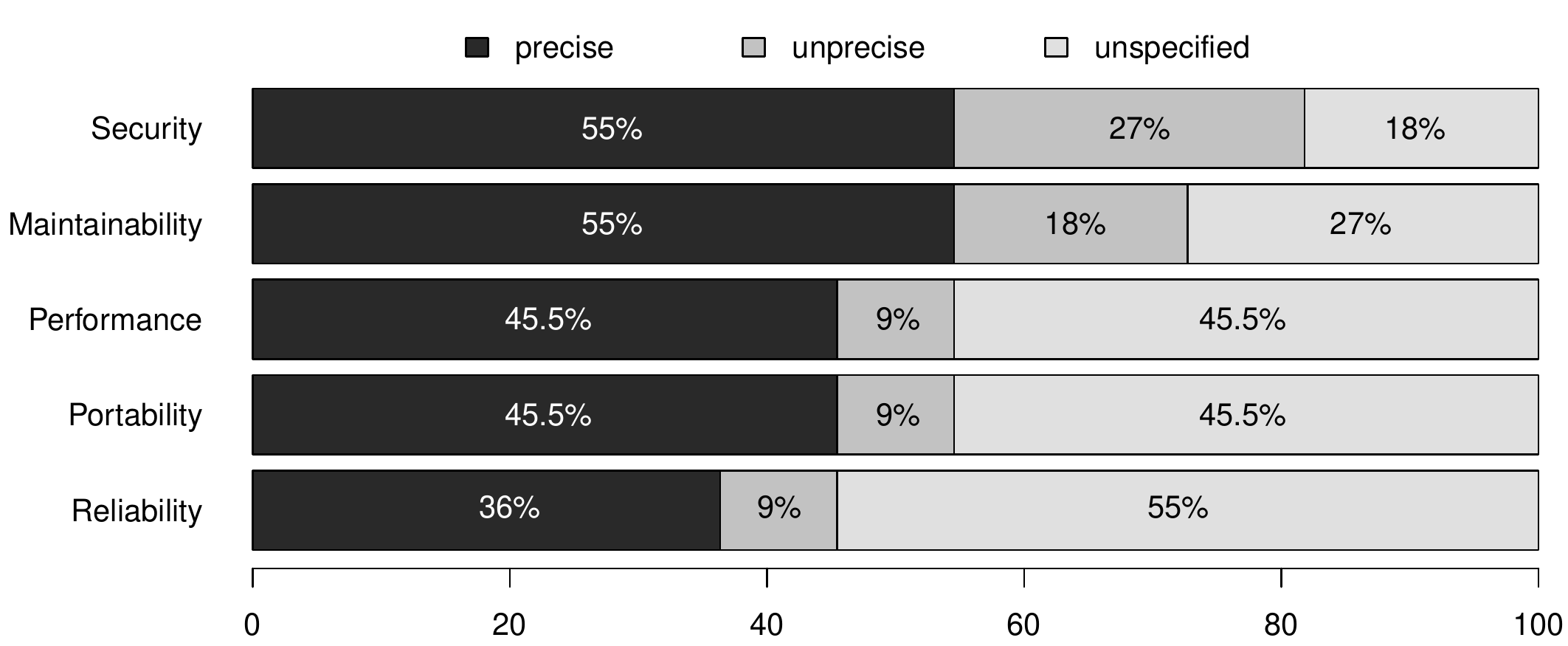}
\caption{Preciseness of NFR specification in the interviewed companies.\label{fig:nfr_specification_quality}}
\label{fig:nfr_preciseness}
\end{figure}

In another question we asked the experts how precisely security, maintainability, performance, portability and reliability requirements are defined in their companies using concrete quantitative measures. As can be depicted from figure \ref{fig:nfr_preciseness}, security and maintainability requirements are most precisely defined in the interviewed companies. The preciseness of performance and portability requirements are a minor interest for these companies. Reliability requirements are mostly not specified at all. The respective experts argues that their companies \textit{``...rely on the middleware to ensure this.''}

It can be summarized that security and maintainability requirements are specified in most interviewed companies. On the other hand, runtime characteristics such as performance, portability, and reliability are specified either unprecise or not specified at all. 

\iffalse Finally, we asked the experts how contradictions between NFRs are detected during requirements engineering in their companies. Interestingly, none of the interviewed companies have a structured methodology in place to help them detecting contradictions in an early phase of the software lifecycle.}{Gibt es Aussagen dazu, ob das aber gewünscht wäre bzw. als wichtig angesehen wird.}\fi

\subsection{Classification of Stakeholder Interests}
\label{interests_classification_schema}
The second research question examines how the different types of NFRs that typically arise in the projects can be condensed into a classification schema for stakeholder interests. The objective hereby is to identify classes of similar interests which are common for specific types of stakeholders (e.g., from the legal, business or engineering domain). This classification schema provides a structure for eliciting the stakeholder interests in the respective step of our method (cf. section \ref{interest_elicitation}).\\

Based on existing taxonomies for NFRs of Chung et al. \cite{chung_non-functional_2009}, Roman \cite{roman_taxonomy_1985}, and Sommerville \cite{sommerville_software_2016} we initially categorized the answers of the experts into 12 classes of stakeholder interests. We only included those classes of stakeholder interests which have at least been mentioned by one of the experts as being relevant. This is due to the experience of the experts that some interests are practically more important than others. 

After an additional refinement step we condensed these stakeholder interests into 9 classes according to the following classification schema:

\begin{itemize}\setlength\itemsep{3pt}
\item \textbf{Interface interests} come in two flavors: 
\begin{itemize}\setlength\itemsep{2pt}
\item	\textit{User interface interests} concern how the system should interact with users. This especially affects the users' interaction device (e.g., mobile apps, desktop applications) and the modality of the interaction (e.g., through virtual or physical keyboards, gesture-based or motion-controlled interactions).
\item	\textit{Application interface interests} address the user's interactions with other systems during execution of a task. These interests especially concern the peculiarities of other systems, for example in terms of network protocols, transaction management, or data formats.
\end{itemize}
\item	\textbf{Informational interests} denote objectives regarding the visual and semantic expression that must be taken into account when presenting information to the user. These interests typically arise when the prospective user groups of the software process information differently, such as e.g., elderly users and digital natives or users of different cultures. Also, these interests address accessibility aspects aiming on the inclusion of people with disabilities.
\item	\textbf{Behavioral interests} describe how the system shall behave under normal and exceptional conditions. These interests primarily address time-based response behavior, resource utilization (e.g., bandwidth or memory utilization), reliability, availability, confidentiality, and ability to cope with catastrophes or unplanned service outages.
\item	\textbf{Operating interests} focus on how the later solution shall be deployed, monitored, and operated. These typically address cloud-service provisioning or virtualization, the use of application servers and operating systems, networks, and access to physical IT systems.
\item	\textbf{Human interests} cover motivational aspects and special skills of users and DevOps teams required for operating the software.
\item	\textbf{Lifecycle interests} address how the software system should align with project management methodologies and engineering standards to assure its maintainability, comprehensibility, and portability throughout its lifecycle.
\item	\textbf{Economic interests} align the software system with the business model, revenue streams, and performance indicators. These interests typically are derived from make-or-buy and product strategy decisions.
\item	\textbf{Data governance interests} delineate how data should be processed and stored among the systems which are involved in task execution in order to facilitate data backtracking, governance, and updating. 
\item	\textbf{Legal and policy interests} address how policies and legal issues, e.g., privacy and copyright liabilities, data export prohibitions (e.g., embargoed countries), and commercial and fiscal laws can be obeyed. They also capture implications of the commercial use of open-source licensed software or source-code ownership restrictions arising from the intellectual property rights of software developers.  
\end{itemize}

\subsection{Task-Oriented Modeling of Functional Requirements and Tailoring of Stakeholder Interests}
In the frame of the third research question we performed an evaluation of our task-oriented modeling and interest tailoring approach through expert interviews in the different companies.

\subsubsection{Applicability of the Task-Oriented Modeling Approach} The interviews included six questions to evaluate the applicability of our task-oriented modeling approach for different objectives, identify benefits, and weaknesses and determine whether the experts would use this approach in their projects. First, we presented each expert a sample set of user tasks and task details for a well-known online book store and described how these concepts would be modeled and interlinked in our approach. Then, we asked them whether the combined display of tasks, information objects, control flows, and implementation endpoints (cf. figure \ref{fig:flow}) would help them to:
\begin{itemize}
\item[$\bullet$] understand the utility of the task for the user; 
\item[$\bullet$] detect missing but critical functionality to achieve the user's goal; and
\item[$\bullet$] prioritize tasks by provided utility, for example to sketch a minimum viable product (MVP).
\end{itemize}

\begin{figure}[h]
\vspace{-10pt}
\centering
\includegraphics[scale=0.6]{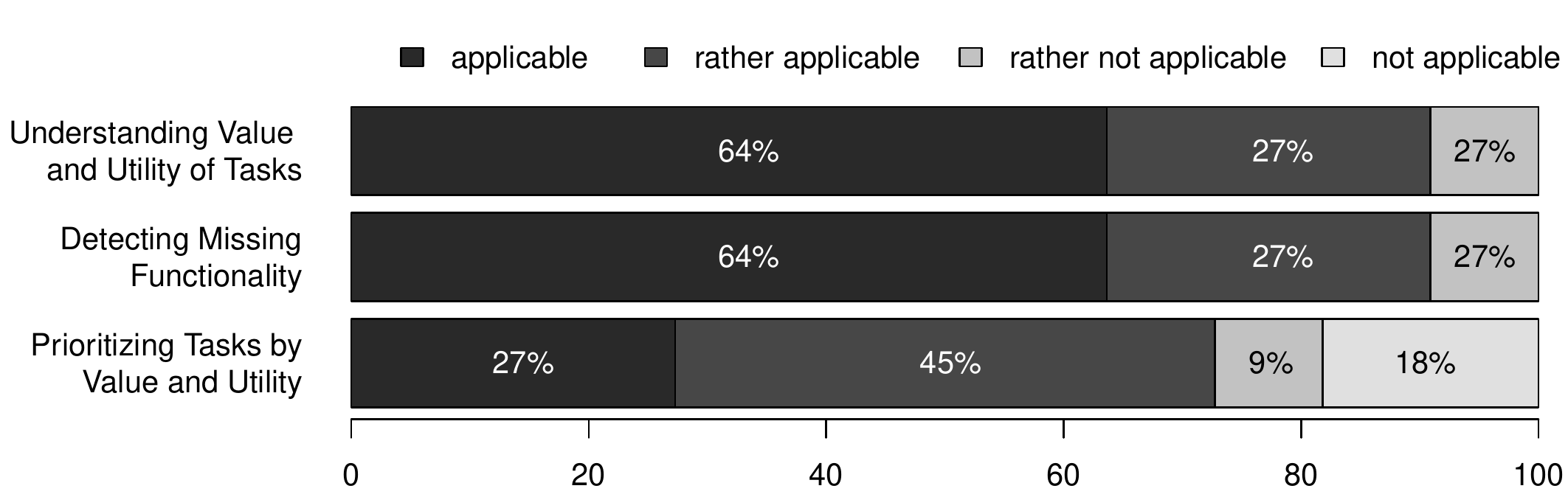}
\caption{Applicability of the task-oriented modeling approach for selected objectives.\label{fig:applicability_task_modeling}}
\end{figure}

As depicted in figure \ref{fig:applicability_task_modeling}, the experts responded positively that our approach helps them to understand the provided utility and goal of each task for the user and also to detect whether critical functionality is missing for the goal to be attained. The decomposition of the superordinate user goal into tasks and subtasks in particular fosters intuitive understanding of what objectives are attained through the user and what is required from an engineering perspective to implement the task. 

While 73\% of the experts rated the approach applicable or rather applicable to prioritize tasks according to the provided utility for the user, 18\% rated it applicable for this purpose. Our approach, however, does not aim to support task prioritization but rather assumes that tasks have already been selected, for example based on the feature set of the MVP. 

The experts were also asked if any additional modeling elements would be necessary for them to use the approach in their projects. Among their answers, most frequently mentioned were the requirements for demarking user roles and labeling tasks with tags along with better support for describing exceptions and the interdependences among tasks and user-interface elements.

Next, we asked the experts to rate our concept of information objects to explicate information that is consumed or generated during execution of a user task. While all experts argued that information objects support them to model task-related information, they see different benefits associated with them. Automated generation of domain objects has been mentioned by 44\% of the experts. They summarized that when adhering to a formal modeling standard such as UML, domain objects can be generated with little effort. Explicating implicit domain knowledge, which is often only poorly documented, has been mentioned as a benefit by 22\% of the experts. A share of 17\% of experts argued that information objects would help them to better understand functional requirements as they always depend on each other. Facilitating the evaluation of implementation feasibility was highlighted by 11\% of the experts. They argued that information objects help them to more easily detect interdepencencies and obstacles between requirements. In the same line of argumentation, 5\% of experts stressed that information objects are specially beneficial to resolve data dependencies between functional requirements.

\begin{figure}[h]
\centering
\includegraphics[scale=0.6]{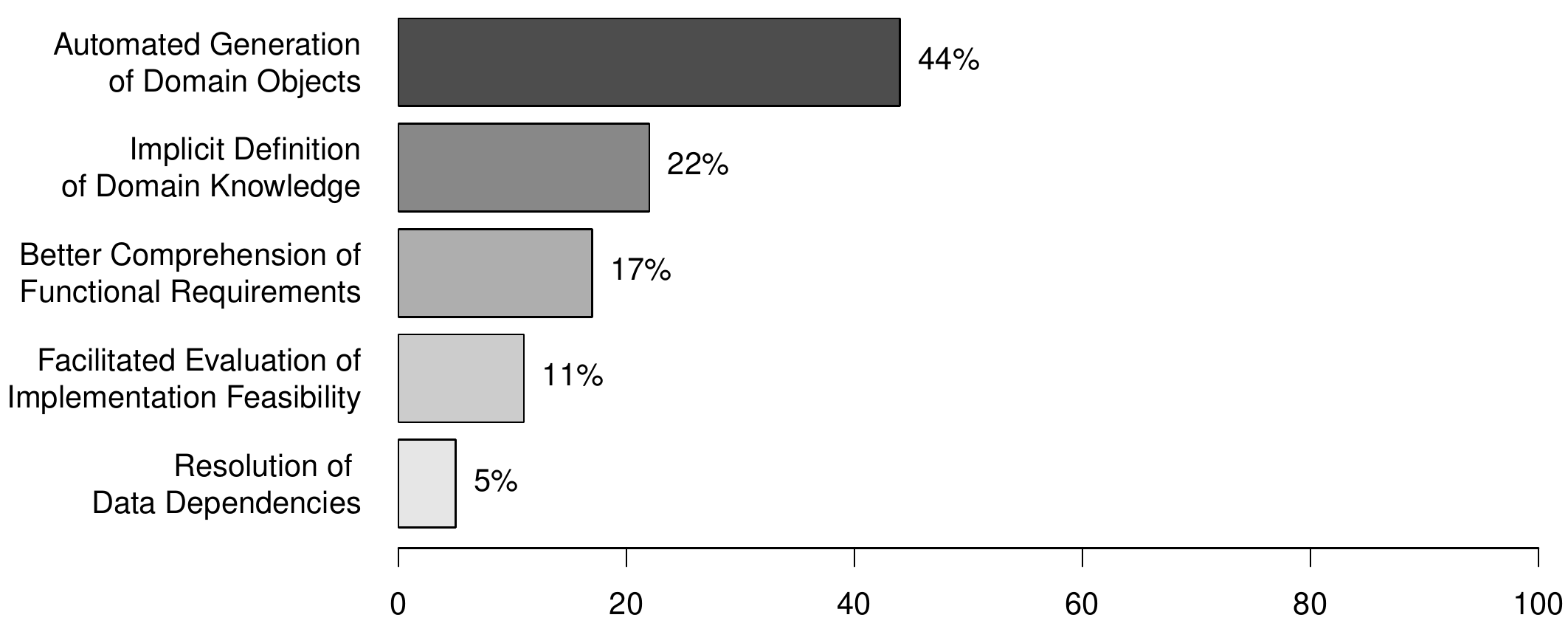}
\caption{Benefits of information objects for task-oriented modeling of functional requirements.\label{fig:information_objects}}
\end{figure}

\subsubsection{Suitability of the Task-Oriented Modeling Approach} Finally, the experts were asked to rate the suitability of the approach for modeling functional requirements, again on a 4-point Likert scale, and to justify their answers. 

\begin{figure}[h]
\centering
\includegraphics[scale=0.6]{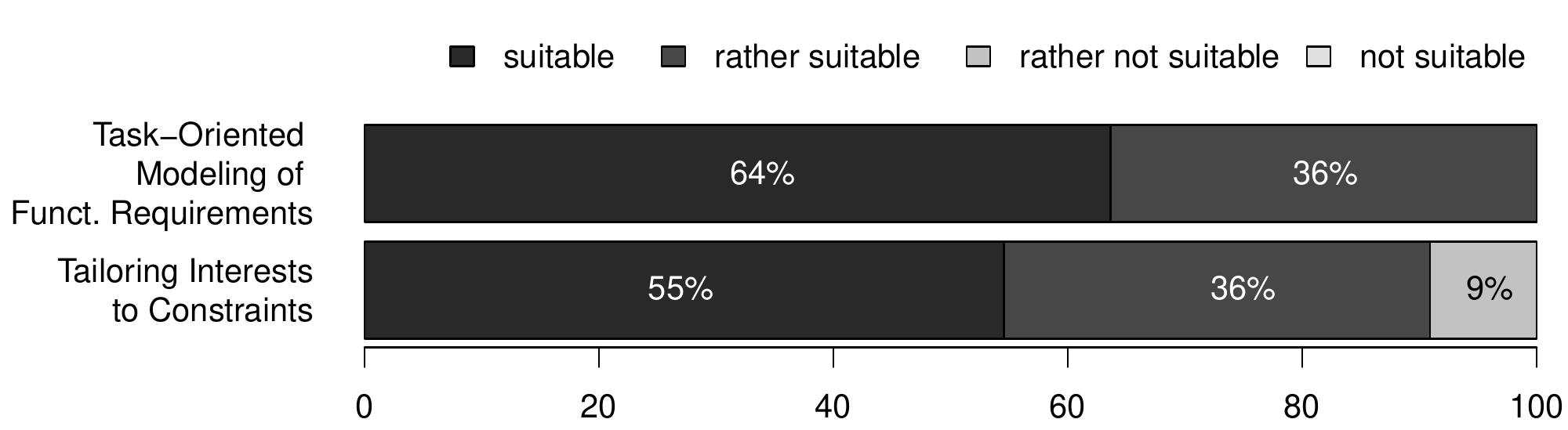}
\caption{Suitability of the task-oriented modeling approach.\label{fig:suitability_task_modeling}}

\end{figure}

Figure \ref{fig:suitability_task_modeling} illustrates that the answers gave a clear picture of the approach's suitability, as all experts positively rated our approach as either suitable (64\% of experts) or rather suitable (36\% of experts). No responses indicated that the approach could not be applied in practice.
 Next, the experts were asked to elaborate the benefits and weaknesses of the approach. They argued that especially due to the granularity and focus on user tasks, the main benefits of the approach include:
\begin{itemize}\setlength\itemsep{2pt}
\item[$\bullet$] fostering shared understanding of functional requirements among stakeholders especially across domains;
\item[$\bullet$] easier and more precise structuring of the functional scope compared to e.g., user stories or epics;
\item[$\bullet$] additional guidance for eliciting functional requirements through the structured decomposition of goals into tasks and subtasks.
\end{itemize}
Concerning the weaknesses of the method, 27\% of the experts recognized no weaknesses for the practical introduction of the method. We codified the open answers of the 73\% experts naming weaknesses, summarizing their statements as follows. 34\% responded that, as with every structured method, our approach requires knowledge of how to use it effectively in projects. The experts also assumed that the method would require only little training time for its practical introduction, due to its simplicity. The additional maintenance effort to adapt diagrams to changing requirements was expressed as a weakness by 22\% of the experts, and the complexity of the model elements was also mentioned by 22\% of the experts. Flexibility and additionally introduced complexity were each mentioned by 11\% of the experts.

Before asking the experts to rate our approach, we presented them a selection of tasks from a well-known online book store and a list of generally understandable performance, privacy, and legal interests (cf. section \ref{interest_elicitation}). Then, we illustrated how these interests can be used in our approach to derive constraints on the level of user tasks. Finally, we asked the experts 4 questions to evaluate our method for tailoring these interests in the context of individual user tasks.

\subsubsection{Suitability of the Interest-Tailoring Approach} The experts were asked to rate the overall suitability of the approach on a 4-point Likert scale. Again we received predominantly positive feedback, with 55\% experts judging the approach as suitable and 36\% as rather suitable. The 9\% of experts judging the approach as rather unsuitable argued that a precise and comprehensive specification of constraints from interests should only be done for selected features and not on the level of user tasks. No expert judged the interest tailoring approach as unsuitable for their projects.

\subsubsection{Benefits and Weaknesses of the Interest-Tailoring Approach.} 
Finally, we asked the experts 2 open questions to elaborate the benefits of our interest tailoring approach on the level of user tasks. We condensed their answers to these questions into 8 categories. Increased comprehensibility of the system was expressed as a benefit by 82\% of the experts, namely by increasing the clarity of objectives pursued through an interest. In 55\% of the interviews, the experts mentioned the increased specification quality of constraints derived from interests, and 36\% said it would help them to assess project risks by better understanding interests and their interdependencies.

In 18\% of interviews, experts expressed the prioritization of interests, the time savings accrued by deriving constraints from interests, and the ease of documentation as benefits of the approach. Only 9\% of experts mentioned that our approach could also help them to detect critical paths.
Based on the open answers examining the weaknesses, we codified the experts' answers into 3 groups. 45\% of experts believed that the approach would introduce an additional specification effort but also expressed that the expected benefits outweighed these tradeoffs accompanying any structured method. 27\% of experts mentioned the complexity of the approach as a drawback, and a further 18\% anticipated that our approach would result in explicitly specifying standard industry constraints that usually need no special documentation (e.g., a default availability, common security requirements).

\section{Threats to Validity}
\label{threats}
We see a threat to \textbf{construct validity}
in the different interpretations of the questions by the experts, which is mainly due to their different roles and experiences. We addressed this threat by showing each expert concrete definitions of the terminology used in the interview and discussed any ambiguities. When summarizing the interview answers, we also considered the background and role of each expert to determine from what view and with what intention the statement was given. Also, as the objective of the interviews conducted in our exploratory case study was on gathering feedback for methodological improvement of the method, a further empirical validation is necessary after the experts have applied our method. 

The foremost threat to \textbf{internal validity}
can be seen in some experts’ trend to answer in confirmation of our theories, which became evident when elaborating on the practice of eliciting constraints. This could have led to confirmation bias, but we regard this as negligible because in response to this trend to answer towards confirming our theories, we asked follow-up questions to capture the experts’ actual experiences.

We addressed the threat to \textbf{external validity} by selecting experts who operate in different industry sectors, and we also selected only one expert per company. However, we see a threat to the generalizability of the results to other industries due to the different size and maturity of requirements engineering practices in the companies. Also,  an empirical validation of the approach needs to be conducted to gather empirical evidence about the generalizability of the results.

\section{Conclusion and Future Work}
\label{conclusion}
Utilizing the proposed hierarchical task analysis approach to structurally decompose tasks into subtasks seems to offer a promising approach to clarify the core functionality needed to support the users’ goals for involved stakeholders. Stakeholder interests can be effectively tailored to constraints within the context of a user task so that each interest can be fulfilled in a satisfying but not inevitably optimal way. Our approach explicitly addresses challenges arising from both the heterogeneity of stakeholder interests and the interdependences between functional requirements and NFRs in software systems. 

We have focused our exploratory case study on the elicitation practice and specification preciseness of NFRs in the companies. The main objective of this research design was to effectively address the issues and challenges gathered from the expert interviews in our method. Together with an industry partner we are currently planning an empirical validation of the approach to ensure the generalizability and the applicability of the results in an industrial context.

Future work shall concentrate on ensuring scalability of the method for large-scale software engineering projects. Also, we are currently defining an operational quality model for specifying the constraints of the task-constraint matrix using concrete measures. Along with suitable instruments that operationalize these measures this allows to automatically monitor fulfillment of stakeholder interests throughout the DevOps cycle.

\renewcommand{\refname}{Bibliography}
\printbibliography

@misc{haindl_se_workhsop,
  title = {{Tailoring} and {Evaluating} {Non-Functional} {Interests} {Towards} {Task-Oriented} {Functional} {Requirements}},
  author = {Haindl, Philipp and Plösch, Reinhold},
  url = {http://ceur-ws.org/Vol-2581/aesp2020paper1.pdf},
  year = {2020},
  booktitle = {{Combined} {Proceedings} of the {Workshops} at {Software} {Engineering} 2020 {Co-located} with the {German} {Software} {Engineering} {Conference} 2020 (SE 2020)},
  note = {accessed 2022/01/15},
  volume = {2581}
}

@article{runeson_guidelines_2009,
	title = {Guidelines for conducting and reporting case study research in software engineering},
	volume = {14},
	issn = {1382-3256, 1573-7616},
	url = {http://link.springer.com/10.1007/s10664-008-9102-8},
	doi = {10.1007/s10664-008-9102-8},
	language = {en},
	number = {2},
	urldate = {2015-02-12},
	journal = {Empirical Software Engineering},
	author = {Runeson, Per and Höst, Martin},
	month = apr,
	year = {2009},
	pages = {131--164}
}

@article{karlsson_requirements_2007,
	title = {Requirements engineering challenges in market-driven software development – {An} interview study with practitioners},
	volume = {49},
	issn = {0950-5849},
	doi = {10.1016/j.infsof.2007.02.008},
	number = {6},
	journal = {Qualitative Software Engineering Research},
	author = {Karlsson, Lena and Dahlstedt, Åsa G. and Regnell, Björn and Natt och Dag, Johan and Persson, Anne},
	month = jun,
	year = {2007},
	pages = {588--604}
}

@incollection{annett_hierarchical_2003,
	address = {London, UK},
	title = {Hierarchical {Task} {Analysis}},
	isbn = {978-0-8058-4433-7},
	booktitle = {The {Handbook} of {Task} {Analysis} for {Human}-{Computer} {Interaction}},
	publisher = {Taylor \& Francis},
	author = {Annett, John},
	year = {2003},
	pages = {67--82}
}

@article{lauesen_task_2012,
	title = {Task descriptions versus use cases},
	volume = {17},
	issn = {1432-010X},
	url = {https://doi.org/10.1007/s00766-011-0140-1},
	number = {1},
	journal = {Requirements Engineering},
	author = {Lauesen, Soren and Kuhail, Mohammad A.},
	month = mar,
	year = {2012},
	pages = {3--18}
}

@inproceedings{yang_linking_2012,
	title = {Linking {Functions} and {Quality} {Attributes} for {Software} {Evolution}},
	volume = {1},
	doi = {10.1109/APSEC.2012.151},
	booktitle = {2012 19th {Asia}-{Pacific} {Software} {Engineering} {Conference}},
	author = {Yang, H. and Zheng, S. and Chu, W. C. C. and Tsai, C. T.},
	month = dec,
	year = {2012},
	pages = {250--259}
}

@article{blaine_software_2008,
	title = {Software {Quality} {Requirements}: {How} to {Balance} {Competing} {Priorities}},
	volume = {25},
	issn = {0740-7459},
	doi = {10.1109/MS.2008.46},
	number = {2},
	journal = {IEEE Software},
	author = {Blaine, J. D. and Cleland-Huang, J.},
	month = mar,
	year = {2008},
	pages = {22--24}
}

@article{ozkaya_making_2008,
	title = {Making {Practical} {Use} of {Quality} {Attribute} {Information}},
	volume = {25},
	issn = {0740-7459},
	doi = {10.1109/MS.2008.39},
	number = {2},
	journal = {IEEE Software},
	author = {Ozkaya, I. and Bass, L. and Sangwan, R. S. and Nord, R. L.},
	month = mar,
	year = {2008},
	pages = {25--33}
}

@article{regnell_supporting_2008,
	title = {Supporting {Roadmapping} of {Quality} {Requirements}},
	volume = {25},
	issn = {0740-7459},
	doi = {10.1109/MS.2008.48},
	number = {2},
	journal = {IEEE Software},
	author = {Regnell, B. and Svensson, R. B. and Olsson, T.},
	month = mar,
	year = {2008},
	pages = {42--47}
}

@inproceedings{fotrousi_quality_2014,
	title = {Quality requirements elicitation based on inquiry of quality-impact relationships},
	doi = {10.1109/RE.2014.6912272},
	booktitle = {2014 {IEEE} 22nd {International} {Requirements} {Engineering} {Conference} ({RE})},
	author = {Fotrousi, F. and Fricker, S. A. and Fiedler, M.},
	month = aug,
	year = {2014},
	pages = {303--312}
}

@incollection{zowghi_requirements_2005,
	address = {Berlin, Heidelberg},
	title = {Requirements {Elicitation}: {A} {Survey} of {Techniques}, {Approaches}, and {Tools}},
	isbn = {978-3-540-28244-0},
	booktitle = {Engineering and {Managing} {Software} {Requirements}},
	publisher = {Springer Berlin Heidelberg},
	author = {Zowghi, Didar and Coulin, Chad},
	editor = {Aurum, Aybüke and Wohlin, Claes},
	year = {2005},
	doi = {10.1007/3-540-28244-0_2},
	pages = {19--46}
}

@incollection{rolland_modeling_2005,
	address = {Berlin, Heidelberg},
	title = {Modeling {Goals} and {Reasoning} with {Them}},
	isbn = {978-3-540-28244-0},
	url = {https://doi.org/10.1007/3-540-28244-0_9},
	booktitle = {Engineering and {Managing} {Software} {Requirements}},
	publisher = {Springer Berlin Heidelberg},
	author = {Rolland, Colette and Salinesi, Camille},
	editor = {Aurum, Aybüke and Wohlin, Claes},
	year = {2005},
	pages = {189--217}
}

@incollection{berander_requirements_2005,
	address = {Berlin, Heidelberg},
	title = {Requirements {Prioritization}},
	isbn = {978-3-540-28244-0},
	booktitle = {Engineering and {Managing} {Software} {Requirements}},
	publisher = {Springer},
	author = {Berander, Patrik and Andrews, Anneliese},
	editor = {Aurum, Aybüke and Wohlin, Claes},
	year = {2005},
	pages = {69--94}
}

@incollection{chung_non-functional_2009,
	address = {Berlin, Heidelberg},
	title = {On {Non}-{Functional} {Requirements} in {Software} {Engineering}},
	isbn = {978-3-642-02462-7},
	booktitle = {c},
	publisher = {Springer},
	author = {Chung, Lawrence and Leite, Julio Cesar Prado},
	year = {2009},
	pages = {363--379}
}

@book{chung_non-functional_2000,
	series = {International {Series} in {Software} {Engineering}},
	title = {Non-{Functional} {Requirements} in {Software} {Engineering}},
	volume = {5},
	isbn = {978-0-7923-8666-7},
	publisher = {Springer US},
	author = {Chung, L. and Nixon, B.A. and Yu, E. and Mylopoulos, J.},
	year = {2000}
}

@inproceedings{cleland-huang_goal-centric_2005,
	title = {Goal-centric traceability for managing non-functional requirements},
	doi = {10.1109/ICSE.2005.1553579},
	booktitle = {Proceedings. 27th {International} {Conference} on {Software} {Engineering}, 2005. {ICSE} 2005.},
	author = {Cleland-Huang, J. and Settimi, R. and BenKhadra, O. and Berezhanskaya, E. and Christina, S.},
	month = may,
	year = {2005},
	pages = {362--371}
}

@inproceedings{lamsweerde_goal-oriented_2001,
	title = {Goal-oriented requirements engineering: a guided tour},
	doi = {10.1109/ISRE.2001.948567},
	booktitle = {Proceedings {Fifth} {IEEE} {International} {Symposium} on {Requirements} {Engineering}},
	author = {Lamsweerde, A. van},
	year = {2001},
	pages = {249--262}
}

@article{broy_rethinking_2015,
	title = {Rethinking {Nonfunctional} {Software} {Requirements}},
	volume = {48},
	issn = {0018-9162},
	doi = {10.1109/MC.2015.139},
	number = {5},
	journal = {Computer},
	author = {Broy, M.},
	month = may,
	year = {2015},
	pages = {96--99}
}

@book{bourque_guide_2014,
	address = {Los Alamitos, CA, USA},
	edition = {3},
	title = {Guide to the {Software} {Engineering} {Body} of {Knowledge}},
	isbn = {978-0-7695-5166-1},
	publisher = {IEEE Computer Society Press},
	author = {Bourque, Pierre and Fairley, Richard E.},
	year = {2014}
}

@inproceedings{glinz_rethinking_2005,
	address = {Munich, Germany},
	title = {Rethinking the notion of non-functional requirements},
	volume = {2},
	booktitle = {Proceedings of the {Third} {World} {Congress} for {Software} {Quality} (3WCSQ 2005)},
	author = {Glinz, Martin},
	year = {2005},
	pages = {55--64}
}

@article{noauthor_iso/iec/ieee_2011-1,
	title = {{ISO}/{IEC}/{IEEE} {International} {Standard} - {Systems} and software engineering – {Life} cycle processes –{Requirements} engineering},
	doi = {10.1109/IEEESTD.2011.6146379},
	journal = {ISO/IEC/IEEE 29148:2011(E)},
	month = dec,
	year = {2011},
	pages = {1--94}
}

@article{noauthor_iso/iec/ieee_2018,
	title = {{ISO}/{IEC}/{IEEE} 25030:2007 {International} {Standard} - {Software} product {Quality} {Requirements} and {Evaluation} ({SQuaRE}) - {Quality} requirements},
	url = {https://www.iso.org/standard/35755.html},
	urldate = {2022-01-15},
	journal = {25030:2007},
	year = {2018},
	note = {(accessed 2022/01/15)}
}

@article{roman_taxonomy_1985,
	title = {A taxonomy of current issues in requirements engineering},
	volume = {18},
	issn = {0018-9162},
	doi = {10.1109/MC.1985.1662861},
	number = {4},
	journal = {Computer},
	author = {{Roman}},
	month = apr,
	year = {1985},
	pages = {14--23}
}

@inproceedings{riegel_systematic_2015,
	title = {A {Systematic} {Literature} {Review} of {Requirements} {Prioritization} {Criteria}},
	url = {https://link-1springer-1com-1um9stgpp0030.han.ubl.jku.at/chapter/10.1007/978-3-319-16101-3_22},
	doi = {10.1007/978-3-319-16101-3_22},
	language = {en},
	urldate = {2018-12-18},
	booktitle = {Requirements {Engineering}: {Foundation} for {Software} {Quality}},
	publisher = {Springer, Cham},
	author = {Riegel, Norman and Doerr, Joerg},
	month = mar,
	year = {2015},
	pages = {300--317}
}

@inproceedings{herrmann_exploring_2007,
	title = {Exploring the {Characteristics} of {NFR} {Methods} – {A} {Dialogue} {About} {Two} {Approaches}},
	url = {https://link-1springer-1com-1um9stgpp0030.han.ubl.jku.at/chapter/10.1007/978-3-540-73031-6_24},
	doi = {10.1007/978-3-540-73031-6_24},
	language = {en},
	urldate = {2018-12-18},
	booktitle = {Requirements {Engineering}: {Foundation} for {Software} {Quality}},
	publisher = {Springer, Berlin, Heidelberg},
	author = {Herrmann, Andrea and Kerkow, Daniel and Doerr, Joerg},
	month = jun,
	year = {2007},
	pages = {320--334}
}

@article{zubcoff_evaluating_2019,
	title = {Evaluating different i*-based approaches for selecting functional requirements while balancing and optimizing non-functional requirements: {A} controlled experiment},
	volume = {106},
	issn = {0950-5849},
	shorttitle = {Evaluating different i*-based approaches for selecting functional requirements while balancing and optimizing non-functional requirements},
	doi = {10.1016/j.infsof.2018.09.004},
	language = {en},
	urldate = {2018-12-28},
	journal = {Information and Software Technology},
	author = {Zubcoff, Jose and Garrigós, Irene and Casteleynb, Sven and Mazón, Jose-Norberto and Aguilar},
	month = feb,
	year = {2019},
	pages = {68--84}
}

@inproceedings{ameller_dealing_2010,
	title = {Dealing with {Non}-{Functional} {Requirements} in {Model}-{Driven} {Development}},
	doi = {10.1109/RE.2010.32},
	booktitle = {2010 18th {IEEE} {International} {Requirements} {Engineering} {Conference}},
	author = {Ameller, D. and Franch, X. and Cabot, J.},
	month = sep,
	year = {2010},
	pages = {189--198}
}

@inproceedings{vierhauser_case_2014,
	address = {New York, NY, USA},
	series = {{ICSE} {Companion} 2014},
	title = {A {Case} {Study} on {Testing}, {Commissioning}, and {Operation} of {Very}-large-scale {Software} {Systems}},
	isbn = {978-1-4503-2768-8},
	url = {http://doi.acm.org/10.1145/2591062.2591179},
	doi = {10.1145/2591062.2591179},
	urldate = {2019-02-14},
	booktitle = {Companion {Proceedings} of the 36th {International} {Conference} on {Software} {Engineering}},
	publisher = {ACM},
	author = {Vierhauser, Michael and Rabiser, Rick and Grünbacher, Paul},
	year = {2014},
	note = {event-place: Hyderabad, India},
	pages = {125--134}
}

@inproceedings{haindl_research_2019,
	address = {Essen, Germany},
	series = {{LNCS} 11412},
	title = {A {Research} {Preview} on {TAICOS} - {Tailoring} {Stakeholder} {Interests} to {Task}-{Oriented} {Functional} {Requirements}},
	isbn = {978-3-030-15537-7},
	doi = {10.1007/978-3-030-15538-4_22},
	booktitle = {Requirements {Engineering}: {Foundation} for {Software} {Quality}},
	publisher = {Springer Nature Switzerland},
	author = {Haindl, Philipp and Plösch, Reinhold and Körner, Christian},
	year = {2019},
	pages = {1--7}
}

@article{ameller_dealing_2019,
	title = {Dealing with {Non}-{Functional} {Requirements} in {Model}-{Driven} {Development}: {A} {Survey}},
	issn = {0098-5589},
	shorttitle = {Dealing with {Non}-{Functional} {Requirements} in {Model}-{Driven} {Development}},
	doi = {10.1109/TSE.2019.2904476},
	journal = {IEEE Transactions on Software Engineering},
	author = {Ameller, D. and Franch, X. and Gómez, C. and Martínez-Fernández, S. and Araujo, J. and Biffl, S. and Cabot, J. and Cortellessa, V. and Méndez, D. and Moreira, A. and Muccini, H. and Vallecillo, A. and Wimmer, M. and Amaral, V. and Bühm, W. and Bruneliere, H. and Burgueño, L. and Goulão, M. and Teufl, S. and Berardinelli, L.},
	year = {2019},
	pages = {1--1}
}

@inproceedings{doerr_non-functional_2005,
	title = {Non-functional requirements in industry - three case studies adopting an experience-based {NFR} method},
	doi = {10.1109/RE.2005.47},
	booktitle = {13th {IEEE} {International} {Conference} on {Requirements} {Engineering} ({RE}'05)},
	author = {Doerr, J. and Kerkow, D. and Koenig, T. and Olsson, T. and Suzuki, T.},
	month = aug,
	year = {2005},
	pages = {373--382}
}

@inproceedings{daneva_software_2013,
	series = {Lecture {Notes} in {Computer} {Science}},
	title = {Software {Architects}’ {Experiences} of {Quality} {Requirements}: {What} {We} {Know} and {What} {We} {Do} {Not} {Know}?},
	isbn = {978-3-642-37422-7},
	shorttitle = {Software {Architects}’ {Experiences} of {Quality} {Requirements}},
	language = {en},
	booktitle = {Requirements {Engineering}: {Foundation} for {Software} {Quality}},
	publisher = {Springer Berlin Heidelberg},
	author = {Daneva, Maya and Buglione, Luigi and Herrmann, Andrea},
	editor = {Doerr, Joerg and Opdahl, Andreas L.},
	year = {2013},
	pages = {1--17}
}

@article{adolph_using_2011,
	title = {Using grounded theory to study the experience of software development},
	volume = {16},
	issn = {1573-7616},
	url = {https://doi.org/10.1007/s10664-010-9152-6},
	doi = {10.1007/s10664-010-9152-6},
	language = {en},
	number = {4},
	urldate = {2019-04-01},
	journal = {Empirical Software Engineering},
	author = {Adolph, Steve and Hall, Wendy and Kruchten, Philippe},
	month = aug,
	year = {2011},
	pages = {487--513}
}

@book{glaser_discovery_2000,
	address = {New Brunswick},
	title = {The {Discovery} of {Grounded} {Theory}: {Strategies} for {Qualitative} {Research}},
	isbn = {978-0-202-30260-7},
	shorttitle = {The {Discovery} of {Grounded} {Theory}},
	language = {English},
	publisher = {Routledge},
	author = {Glaser, Barney and Strauss, Anselm},
	month = jan,
	year = {2000}
}

@book{yin_case_2017,
	address = {Los Angeles},
	edition = {Sixth},
	title = {Case {Study} {Research} and {Applications}: {Design} and {Methods}},
	isbn = {978-1-5063-3616-9},
	shorttitle = {Case {Study} {Research} and {Applications}},
	language = {English},
	publisher = {SAGE Publications, Inc},
	author = {Yin, Robert K.},
	month = nov,
	year = {2017}
}

@book{sommerville_software_2016,
	edition = {10},
	title = {Software {Engineering}},
	isbn = {978-0-13-394303-0},
	urldate = {2019-04-02},
	publisher = {Pearson},
	author = {Sommerville, Ian},
	year = {2016}
}

@inproceedings{berntsson_svensson_quality_2009,
	series = {Lecture {Notes} in {Computer} {Science}},
	title = {Quality {Requirements} in {Practice}: {An} {Interview} {Study} in {Requirements} {Engineering} for {Embedded} {Systems}},
	isbn = {978-3-642-02050-6},
	shorttitle = {Quality {Requirements} in {Practice}},
	language = {en},
	booktitle = {Requirements {Engineering}: {Foundation} for {Software} {Quality}},
	publisher = {Springer Berlin Heidelberg},
	author = {Berntsson Svensson, Richard and Gorschek, Tony and Regnell, Björn},
	editor = {Glinz, Martin and Heymans, Patrick},
	year = {2009},
	pages = {218--232}
}

@misc{prod_dev_patent,
  title={Computer-{Implemented} {Product} {Development} {Method}},
  author={Gehmeyr, A. and Höfler, W. and Kochseder, R. and Rettner, J. and Horn, S.},
  year={2018},
  note={{US} Pat. No. 15/661,498}
}

\end{document}